\documentclass[doublecol,linenumbers]{epl2}

\usepackage[utf8]{inputenc} 
\usepackage{enumerate}
\usepackage{etex}
\reserveinserts{30}
\usepackage{morefloats}

\usepackage{graphicx} 
\usepackage{color}
\usepackage{listings}
\usepackage{afterpage}

\title{Scale-free networks as an epiphenomenon of memory}
\shorttitle{Scale-free networks as an epiphenomenon of memory}
\author{F. Caravelli\inst{1} \and A. Hamma\inst{2} \and M. Di Ventra \inst{3}}
\shortauthor{F. Caravelli \etal}

\institute{
 \inst{1} Department of Computer Science, University College London, Gower Street, London WC1E 6BT, UK and\\
 Invenia Technical Computing, 135 Innovation Dr., Winnipeg, MB R3T 6A8, Canada and\\
 London Institute of Mathematical Sciences, 35a South Street, London W1K 2XF, UK\\
\inst{2} Center for Quantum Information, Institute for Interdisciplinary Information Sciences, Tsinghua University, Beijing 100084, P.R. China \\
\inst{3} Department of Physics, University of California-San Diego, La Jolla, CA 92093, USA
}


\abstract{
Many realistic networks are scale-free, with small characteristic path lengths, high clustering, and power law in their degree distribution. They can be obtained by dynamical networks in which a preferential attachment process takes place. 
However, this mechanism is non-local, in the sense that it requires knowledge of the whole graph in order for the graph to be updated. Instead, if preferential attachment and realistic networks occur in physical systems, these features need to emerge from a local model.
In this paper, we propose a local model and show that a possible ingredient (which is often underrated)
for obtaining scale-free networks with local rules is memory.
Such a model can  be realised in solid-state circuits, using non-linear passive elements with memory such as memristors, and thus can be tested experimentally. }

\begin{document}
\maketitle

The field of complex networks has recently become of tremendous  interest, since the discovery that many --although not
 all-- realistic networks present small-world\cite{ws} and scale-free properties \cite{barev}, namely a power-law tail for the distribution of the degree in the graph.
Scale-free networks are important in a range of topics from ecology and evolution theory\cite{cneco}, to protein folding and neural networks\cite{protcn,protcn2,braincn}, technological and scientific networks \cite{teccn}, social sciences \cite{freeman}, economics \cite{econcomp,econcomp2} and cascade analysis \cite{motterlai}.

  Preferential attachment\cite{barev} is the most well known mechanism for constructing scale-free networks. This is an evolutionary algorithm  in which nodes are added to the network and linked to the existing nodes with a probability proportional to the degree of the pre-existing node. This is a  ``rich gets richer" mechanism that requires at each step the knowledge of the {\it whole} configuration of the graph.
 However, if real-world scale-free networks are created by physical processes, then preferential attachment
must emerge from  completely  local rules.

Several models in which preferential attachment is an emergent
property have been proposed in the physics literature\cite{vazquez}, while other models which made the growth deterministic have been proposed (for instance, \cite{fractalweb}). In particular, the so-called  random walk attachment graph mechanism was proposed
in \cite{sk}, in which new nodes attach to particles hopping on the graph. Other models in which particles act as ``mediators''
for the preferential attachment were proposed in \cite{es,sk,kleinberg,gfwn,cc,ikeda}.
In the case of \cite{sk} and \cite{es}, the links from new nodes are connected to the node in which the particle is sitting after $l$ steps,
but for $l$ big enough, this new link is connected with probability proportional to the degree of the node.  A preferential attachment model related to the weight of the edges was also considered in \cite{bianconire}, without walkers; in \cite{kleinberg}, a model of community formation which generates self-similar graphs was introduced in order to reproduce the distribution of the web structure.

In this article, we present a mechanism that gives rise to scale-free graphs by means of what we call {\it memory}. In the context of this paper, memory is the process given by the interplay between random growth of the graph, decay of the links and their strengthening carried by random walkers that hop over them. We stress that all three processes are entirely local, in the sense that the dynamical processes involving nodes, links and walkers results solely from the interaction of a small number of them in the same space-temporal neighborhood. The temporal  aspect derives from the fact that in order to update some links at the step $t$, we need to know their configuration in that neighborhood at the step $t-k$. We will see that we only need short-term memory ($k=1$) in order to obtain scale-free networks. Such short memory effects are quite general in non-linear systems \cite{reviewdv}.  In fact, {\it any} real condensed-matter system shows some degree of memory in its response functions (e.g., its resistance) when subject to external perturbations \cite{memsys}.

We recall that scale free networks are defined by  the distribution $P(k)$ of the degrees of connectivity, which  obeys a power law  $P(k\gg1)\approx k^{-\rho}$, with
$k$ the number of connections of a given node to neighboring ones, and $\rho$ typically ranging between 2 and 3  \cite{intmath}. Moreover, quite often realistic networks are also "small world", which means that they possess a small average distance between nodes and high clustering coefficient\cite{barev,caldarelli,dyncn,newman,estrada}. The mechanism proposed here produces networks that are both scale free and small world.

Despite the wealth of examples found in the literature, it is still not completely clear how realistic graphs
acquire their scale-free properties through a physical process. Watts and Strogatz demonstrated that small-world networks can be obtained from random networks by adding a few long-range shortcut edges, which connect otherwise distant nodes \cite{ws}. 


As we demonstrate in the following, memory, i.e., the interplay and competition between growth, decay, and strengthening operated by random walkers,  can lead to scale-free, small-world networks. The growth of the network is
the familiar random graph growth. However, as we have anticipated, unlike preferential attachment, the mechanism we  propose is
completely local, i.e, no global information about the graph is needed. We thus suggest that scale-free networks can thus emerge from local {\it self-organization  assisted by memory}.
A similar memory mechanism is used as an optimization procedure
by ants in order to find the shortest path, by reinforcing with pheromones the most walked paths \cite{ants, ants2}. This mechanism is also
the same one employed by networks of memristors (resistors with memory) to solve optimization problems such
as the maze \cite{paralleldv} or other  shortest-path problems \cite{shortpdv}. These memristive networks can support self-organized critical
states \cite{crit1} similar to those encountered in the brain at rest \cite{crit2}. Our predictions can then be readily tested in these types of condensed-matter systems,
and may be relevant to brain dynamics and neurogenesis.\\\ \\

{\em Model.---} The algorithm to create and update the network consists of the following four steps:\\

{\em Initialization:} Start with a weighted random graph with $N_0$ nodes,  link strengths taking random values within [0,1], and $P \leq N_0$ particles placed at random on the nodes. After initialization,  a cycle of the algorithm consists of the steps of Hopping, Strenghtening/Decay effect, and Growth:\\

{\em Hopping:} Let the particles hop between nodes $i$ and $j$ with probability $p_{ij}$ proportional to the link strength $ p_{ij}={A_{ij}}/{\sum_j A_{ij}}$, where $A_{ij}$ is the weighted adjacency matrix of the graph.\\

{\em Strenghtening/Decay:} All the links hopped on by the particles in the last $M$ steps are reinforced by $\gamma$. Links with strength less than threshold $L_d$ decrease their strength of a value $\alpha$, with probability $p_d$, and are removed when they reach a negative weight.\\

{\em Growth:} At this step, a new node is added (and with probability $p_p$ a new particle is placed on it). The new node connects to each of the existing nodes with probability $p_{nl}$ and with random strength between $0$ and $1$ with flat probability distribution.

The simulation stops when $N_f$ nodes are reached. As one can see, the reinforcing process due to the particles hopping is the only mechanism preventing the graph from being eroded. Note  that in \cite{ikeda}, Ikeda introduced a model of reinforcement-decay that bears some similarities with the one introduced in this paper. However, that model features a fixed number of nodes and a initial fixed geometry and dimensionality, and is focused more on the relation between initial topology and diffusion than on the creation of scale-free networks. Note also that the requirement of an initial lattice with fixed dimensionality is very strong. In this paper instead, we are interested in showing how scale-free, small-world networks can arise by means of {\em microscopic rules only}, without any other constraint on the global geometry of the system. As we shall see, in order to obtain the fat tailed distribution of the node degrees without a pre-existing fixed skeleton of geometry, we need a growth mechanism.

{\em Analytical results for some limit cases} .---
Two limit cases can be solved analytically: those in which particles are not present $(P=p_p=0$) and the one in which also decay is not present $(p_d=0$). In the first case, one expects that the nodes of high degree are those whose decay is slower, because they are more likely to have some links above threshold. So the probability of decay would scale like $1/k$, which would yield a corresponding tail in the degree distribution. However, such a graph would not be stable and at long times it would be very sparse. 

In the Supplementary Material of \cite{SupplOurs} we derive and solve a master equation, in mean field theory, for the average degree as a function of time. The master equation takes the form:

\begin{equation}
\partial_t k_s(t)=c - a k_s(t)
\end{equation}
with boundary condition $k_s(s)= c s$. Using then the standard machinery used in the mean field theory analysis, one can calculate an asymptotic distribution which takes the form:

\begin{equation}
P(k\gg1,t\gg1)\approx \frac{1}{t} \cdot \frac{1}{k}.
\end{equation}

We note that the distribution is asymptotically unstable, and indeed a factor $1/t$ is present. This behavior has been confirmed also numerically (shown in the Supplementary Material in \cite{SupplOurs}), and shows that without particles these distributions cannot be stable, as the power law decays with an exponent smaller than 2.

The second case without decay, i.e., $p_{decay}=0$, has been instead discussed in \cite{accnet} where it has been shown that the distribution is well approximated by a Poisson distribution. The other extreme case is the one of $p_{d}=1$ and $\alpha>1$. In this case, the threshold guarantees that only links which are greater than the threshold survive, and thus the effect is similar to reducing the constant $c$.

\begin{figure}
\centering
\includegraphics[scale=0.38]{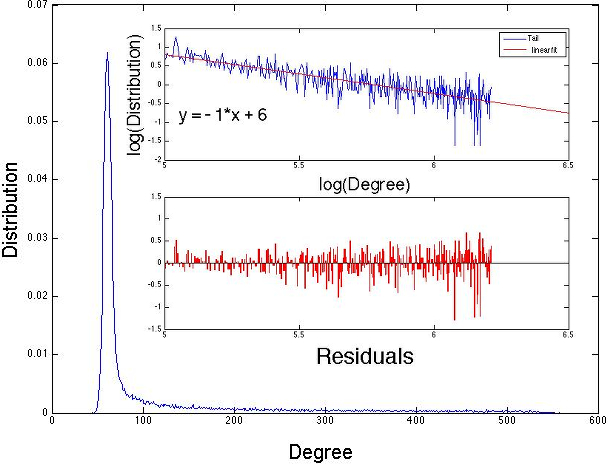}
\caption{Frequency distribution obtained for $p_{nl}=M=1$, $\alpha=p_d=0.01$ and no particles. We observe that the tail follows an approximate power law with exponent $\rho=1$, as predicted by solving the approximate master equation approach. Results averaged over 30 simulation runs.}
\label{fig:onlydecay}
\end{figure}
The power law behavior is consistent with the results obtained in Fig. \ref{fig:onlydecay}, meanwhile the instability of the distribution is supported by the numerical analysis provided in the Supporting Material in \cite{SupplOurs}.

{\em Simulations with particles: results---}
Although a power law is already present in the case of growing graphs with decay, the case without particle leads to a graph which, asymptotically, disappears. Introducing reinforcing particles thus is a necessary requirement to stabilize the graph.
 The simulations were run with a maximum number $N_f$ of 2800 nodes and 2800 particles, starting from a single node with one particle. The decay probability was initially set to  $p_d=0.01$, $p_p=0.5$ and $\alpha=\gamma=0.1$. The threshold parameter for the decay was set to $L_d=0.99$. New nodes were linked to all the old ones with link strength picked at random in the interval $[0,1]$ from a flat distribution.

\begin{figure}
\includegraphics[scale=0.4]{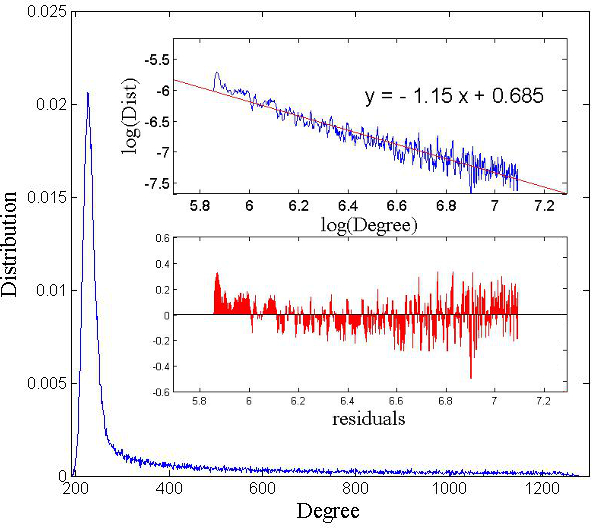}
\caption{Degree distribution and the tail best fit for the parameters $L_d=0.99$,$\alpha=\gamma=0.1$, $p_p=0.5$, $M=1$ and $p_d=0.01$.
We fit the tail using the function $f(d)=a\ d^\rho$. The best fit exponent parameter (using least square fitting) found is $\rho=-1.15\pm 0.4$,  with goodness of fit RMSE$=0.22$ and with adjusted $R^2=0.69$.
The distributin has been obtained after averaging over 30 runs, and has been smoothed using a robust local regression algorithm.}
\label{fig:fit}
\end{figure}

\begin{figure}
\includegraphics[scale=0.4]{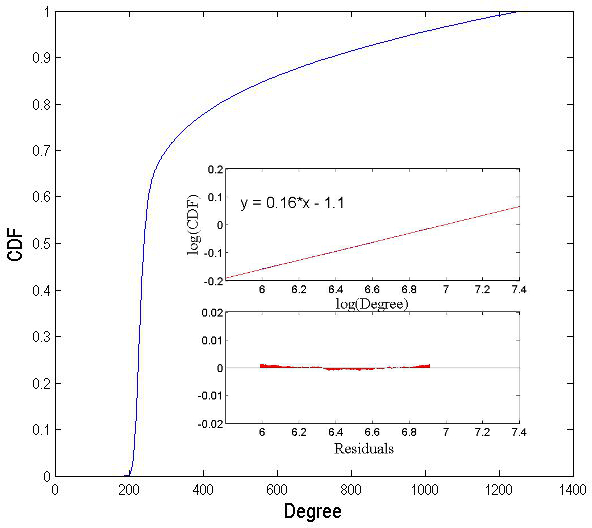}
\caption{Cumulative distribution function of the distribution of Fig. \ref{fig:fit}.  \textit{Inset plot.} Fit of the log-log corresponding to the tail of the distribution on the tail. The log-log plot in the tail of the distribution is linear, with fitted slope $0.16$, compatible with the exponent of the tail of the distribution.}
\label{fig:fitcdf}
\end{figure}

In order to better analyze the properties of these graphs results have been averaged over $30$ runs for fitting the degree distribution, and $20$ for the clustering coefficient. Fig. \ref{fig:fit} shows the results of  our simulations where a power law with exponent $\rho \simeq -1.15$ fits the tail of the degree distribution
 (see also inset of Fig.\ref{fig:fit}) and the fit of the cumulative distribution function in Fig. \ref{fig:fitcdf}. By employing the same parameters but with $p_p=1$ leads to a shorter tail, fitted with an exponent $\rho=-2.36$, roughly double that of the one obtained in Fig. \ref{fig:fit}.
 Since the introduction of particles can lead to tails which fall off with an exponent greater than $\rho=2$, we can interpret this as the fact that particles indeed can lead to stable distributions, as these are now normalizable.

 We analized also the sensitivity of the tail to the size of the system, which in our case is the number of steps of the simulation. With increasing size, the tail becomes longer, and better statistics can be obtained. The current analysis was the best we could obtain with our computing capabilities.

The graphs thus obtained have degree distribution tail exponents both greater and smaller than 2.
This implies a short graph diameter in the latter case, namely an ultra-small network, as guaranteed by the theorem in \cite{theorem}.
To confirm that we have indeed obtained small-world networks we have also studied the clustering coefficient, which is provided in the Supporting Material.

All these results show that the (ant-inspired) memory mechanism is indeed a selection one: decay is a hostile environment which selects those links that are stronger (busy-gets-busier),
by virtue of being crossed more often, which means that there are more roads to them. This competition mechanism does modify the effective exponent of the tail of the distribution, which otherwise would be an unstable distribution.

\begin{figure}

\includegraphics[scale=0.34]{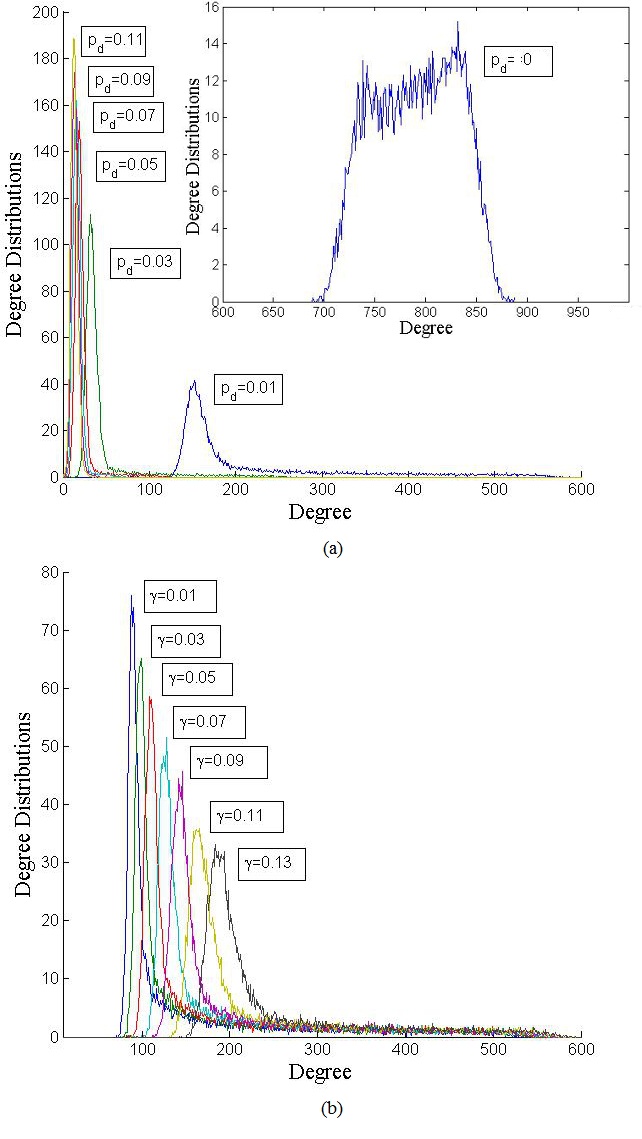}


\caption{{\bf (a)} Frequency distribution of the degree for various values of decay probability $p_d$ , for fixed parameters $L_d=0.99$, $\alpha=\gamma=0.1$, $M=p_p=p_{nl}=1$, and averaged over 30 runs. \textit{ Inset plot.} Degree distribution at $p_d=0$, showing that without decay the fat tailed distribution is not obtained. {\bf (b)} Frequency distribution of degree for various values of $\gamma$, $L_d=0.99$,$\alpha=0.1$, $M=p_{nl}=p_p=1$, $p_d=0.01$ and averaged over 30 simulation runs.}
\label{fig:changeparam}
\end{figure}

In order to confirm this, we have also varied
the decay probability $p_d$ and the reinforcement parameter $\gamma$, by keeping the length $M$ fixed.
This is shown in Fig. \ref{fig:changeparam}(a) for a varying decay probability and in Fig.\ref{fig:changeparam}(b) for a varying reinforcement parameter. In both cases we see that by making the memory too
strong or too weak, the scale-free property is considerably reduced.
For instance in Fig. \ref{fig:changeparam}(a) we see that as the decay probability increases while
 keeping the other parameters fixed, the distribution is skewed towards smaller average degrees. The tails
 of the distribution become shorter and shorter, until eventually they disappear. In the opposite limit, if we switch off the decay mechanism, the scale-free property is completely lost (inset in \ref{fig:changeparam}(a)).

Thus, the introduction of particles, combined with the effect of network acceleration and decay, interpolates between Poisson distributions and an unstable power law with exponent equal to minus one.

This shows that memory, although essential, must decay faster than the time-scale necessary to build the
graph, otherwise a sort of ``memory saturation'' effect occurs that is rather an hindrance to the formation of a
scale-free state. A similar effect holds in networks of memristors~\cite{paralleldv,shortpdv}, where an
optimal memory range is necessary to solve optimization problems. Finally, to make the analogy with the
ant colony even stronger, if the pheromone trail decays too fast---compared to the average time it takes
the ants to go from the nest to the food source---the ants have no time to reinforce the shortest path; if
it does not decay at all, any path is equally attractive to  the next ants, and no efficient
optimization can be achieved.



\begin{figure}
\includegraphics[scale=0.36]{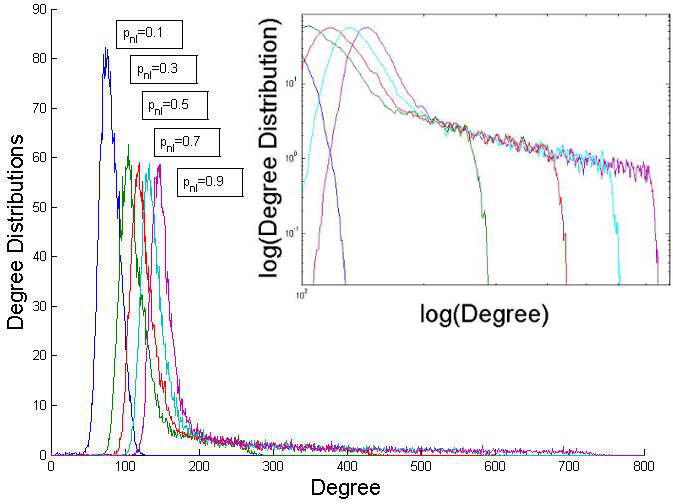}
\caption{Frequency distributions of degree for various values of the parameter $p_{nl}$ for fixed parameters $L_d=0.99$,$\alpha=0.1$, $M=p_p=p_{nl}=1$, $p_d=0.01$ and averaged over 30 simulation runs and smoothed using a robust local regression algorithm. The insert instead represents the double-log of this quantity. The importance of this plot is to show that the exponent of the tail \textit{does not depend on this parameter}, but that the the extension of the tail depends on it, and if fact the closer it is to one, the longer is the tail.}
\label{fig:changedynlinkprob}
\end{figure}

\begin{figure}
\includegraphics[scale=0.36]{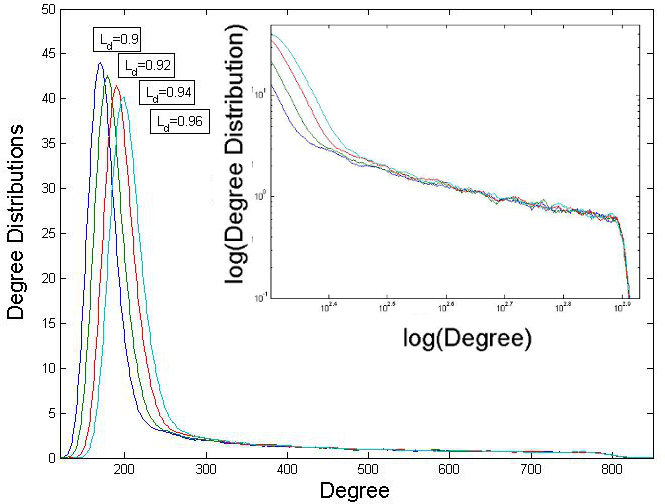}
\caption{Frequency distributions of degree for various values of the stability threshold parameter $L_d$, with $p_{nl}=1$,$\alpha=0.1$, $M=p_{nl}=p_p=1$, $p_d=0.01$; results obtained after  averaging over 30 simulation runs and smoothed using a robust local regression algorithm. The insert instead represents the double-log of this quantity. The importance of this plot is to show that the exponent of the tail is unaffected by the threshold variation, although the peak of the distribution moves to the right for increasing values of the threshold.}
\label{fig:threshvar}
\end{figure}

\begin{figure}
\includegraphics[scale=0.4]{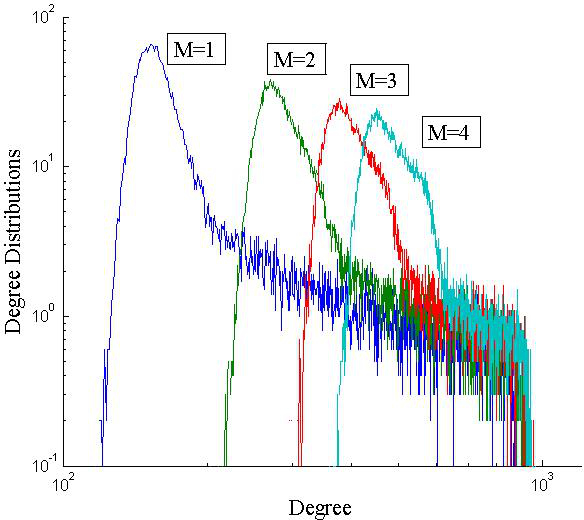}
\caption{Frequency distribution of degree for various values of memory length $M=1,2,3,4$. The results have been obtained with the parameters $\alpha=\gamma=0.1$, $p_{nl}=1$, $p_d=0.01$, $p_b=0.5$ and $L_d=0.99$, and averaged over 30 simulation runs. We see that the tail of the distribution becomes shorter for longer memory indicating an
``optimal memory range'' for scale-free properties. }
\label{fig:changememory}
\end{figure}

In order to stress even more the role of memory in the emergence of scale-free, small-world networks,  we have studied how the length $M$  of memory affects the graph's growth. We find that by increasing $M$ amounts to
softening the selection process since even farther neighbours of high-degree nodes can be reinforced. We show  the results for memory lengths
$M=1,2,3,4$ in Fig. \ref{fig:changememory}.

It is interesting to note that when the memory
length increases, the size of the tail decreases until it actually disappears, indicating that an
``optimal memory range" is necessary for scale-free properties.

{\em Conclusions.---} In this paper we have presented and analyzed a  model of network growth in which scale-free properties emerge by means of a local self-organising mechanism that is based on short-term memory. By this, we mean that as the network grows randomly, all links decay except those that are visited by random walkers, a process that instead increases the link strength. In this model, no previous geometry and dimensionality is assumed, and all of the properties are emergent from the competition of local processes.
The model is inspired by evaporating ant pheromone trails, a process known to be able to solve problems as finding the shortest path between their nest and food by leaving a pheromone track that has a characteristic decay time, but which is reinforced every time other ants use it. In our model the ants are the random walkers.
It turns out that the optimal memory to obtain strong power-law effect must be short-term, but nonzero. Therefore, there is an optimal range of memory length which allows for the emergence of a scaling behavior.

We want to stress also that, being completely local, the model proposed here lends itself to being engineered in the lab. Indeed, this model can be realized in a network of memristive elements (resistors with memory), making our predictions easily realizable experimentally. In much more general terms, our study makes a connection between self-organization, time non-locality and scale-free properties. Since self-organized critical states are ubiquitous in Nature, an interesting line of research suggested by
our work regards the role of memory in the formation of such critical states. We thus hope our work will motivate further theoretical and experimental studies along these directions.
In future works we will address the study of the phase space of the model, in which we observe both fat and short tails, and work on an analytical treatment for the distribution of degree in the case with particles.

\acknowledgments
 This work was supported in part by the National Basic Research Program of China Grant 2011CBA00300, 2011CBA00301, the National Natural Science Foundation of China Grant 61033001, 61361136003. M.D. acknowledges support from the Center for Magnetic Recording Research at UCSD.

\newpage
\section{Supplementary Material}
\begin{figure}
\centering
\includegraphics[scale=0.165]{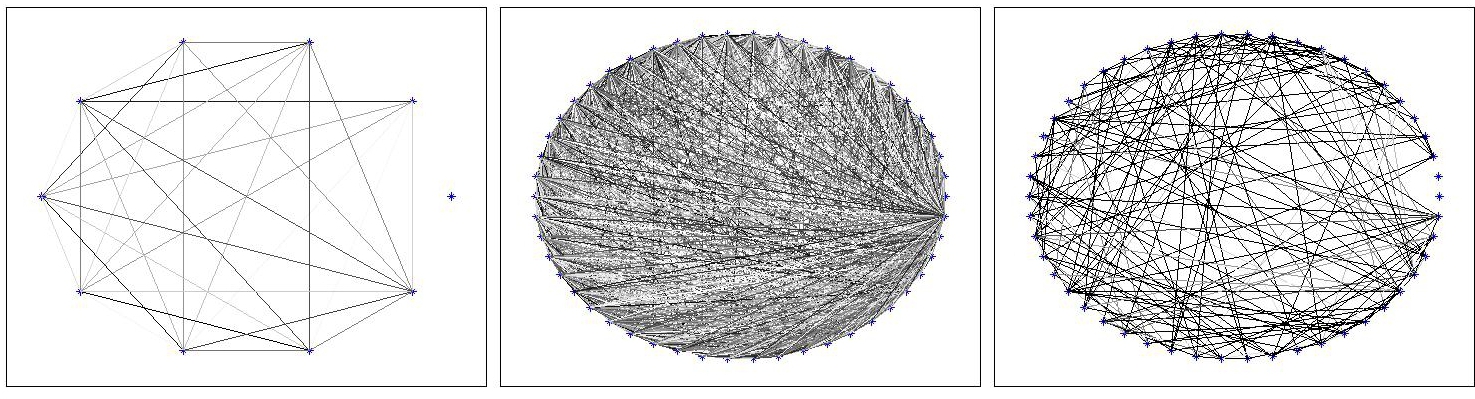}
\caption{Snapshots from the time evolution of the graphs, for parameters $M=p_{nl}=1$,  $L_d=0.99$,$\alpha=\gamma=0.1$, $p_p=0.5$ and $p_d=0.1$. Initial configuration at time T=1 with a single node.
\bf Left: Configuration at time T=10 with 10 nodes.
\bf Center: Configuration at time T=50 with 50 nodes.
\bf Right: Configuration at time T=200 with 50 nodes, after having stopped growth at T=50. We see from this picture that only few links survive the decaying process.}
\label{fig:movie1}
\end{figure}

In order to visualize the difference between the two behaviors, we include two  snapshots from the full dynamics for the graph with memory in Fig. \ref{fig:movie1}.

\textit{Unstable power law without particles.--}  In this section we discuss the master equation for the growth of the graph without particles. Let us start with the dynamic equation for the
 adjacency matrix, which takes in account the decay $\alpha$ and where $\theta$ takes care of the threshold chosen as $L_d=1$:
\begin{eqnarray}
 A^{t+1}_{ij}&=-p_{d} \alpha\ \theta(A^t_{ij})+A_{ij}^t
\end{eqnarray}
the $\theta$ functions instead represent the threshold of the process.

We now introduce some quantities that are useful in calculation of mean properties of the graph for growing networks.
First, we introduce the $p(k,s,t)$, the probability that a vertex introduced at time $s$ has degree $k$ at time $t\ge s$.  At each time step, a new vertex is introduced with probability one. The degree distribution at time $t$, is then given by
\begin{equation}
P(k,t)=\frac{1}{t+1} \sum_{s=0}^t p(k,s,t)
\end{equation}
assuming that we start with one node. The average degree value is given by $k_s(t)=\sum_{k=1}^\infty k\ p(k,s,t)$. This is the degree of the node $s$ at time $t$. Note that $s$ is also a time, as we assume we introduce the node $s$ at time $s$. We use at this point the effective medium approach \textit{Ansatz} \cite{dyncn}, and thus assume that $p(k,s,t)=\delta(k-k_s(t))$. In this approximation, we are able to calculate $P(k,t)$ from the knowledge of $k_s(t)$, using the continuum degree assumption:
\begin{eqnarray}
P(k,t)&=&\frac{1}{t+1} \int_{s=0}^t \delta(k-k_s(t)) ds \nonumber \\
&=& - \frac{1}{1+t} (\frac{\partial k_s(t)}{\partial s})^{-1} |_{s=s(k,t)}.
\label{ansatz}
\end{eqnarray}
One has also to consider the boundary condition $k_t(t)=c t$, given by the fact that the average degree of the new node is proportional to the number of nodes present at time $t$; $c$ is the probability of adding a link to any node in the graph and will play a role in the boundary condition.
One can connect the mean field equation to the dynamical equation for the adjacency matrix using the definition
$k_s(t)=\sum_r A_{sr} (t)$.
  We assume also a continuous time at this point:
\begin{eqnarray}
\frac{d}{dt} A_{ij}(t)&=& -p_{decay}\theta(A_{ij}(t))
\end{eqnarray}
together with the boundary conditions as before. We thus have:
\begin{eqnarray}
\frac{d}{dt} k_s(t) &=& \sum_j \frac{d}{dt} A_{sj}(t) \nonumber \\
&=& -p_{decay}\alpha \sum_j \theta(A_{ij}(t))  \nonumber
\end{eqnarray}
Since we set $L_d=1$, we can approximate, making an error of the same order of magnitude, $\theta(A_{sj}(t))$ with $A_{sj}$. This allows to keep track of the fact that the decay occurs only when a link is present, and allows to have a closed form master equation:
\begin{equation}
\frac{d}{dt} k_s(t)= -p_{decay}\alpha k_s(t)
\end{equation}
In addition to the decay, one has to add a term that represents the growth of the graph:
\begin{equation}
\frac{d}{dt} k_s(t)=c -p_{decay}\alpha k_s(t)
\end{equation}
which according to the definitions of \cite{accnet} is a graph with acceleration.
If at time $t$ we add $t$ links with probability $c$, the average degree is given by $c t$; $c$ can also be interpreted as composed in this effective approach: we add the link with probability $p_l$ and with strength $l_n$. Then $c=p_{nl} \xi$, with $\xi$ the average strength in the mean field theory.

If we now set $p_{decay} \alpha=a$, this equation has a unique solution given by:
\begin{equation}
k(t)=\frac{c}{a}-e^{-a t} Q
\end{equation}
where now $Q$ is a free parameter that we set to $c e^{a s}$ from the boundary condition $k_s(s)=c s$, which implies an equation of the form
\begin{equation}
 k_s(t)=\frac{c \left((a s-1) e^{a (s-t)}+1\right)}{a}
\end{equation}
We now evaluate $s(k,t)$, given by the solution of the implicit equation $k=k_s(t)$:
\begin{equation}
s(k,t)=\frac{W\left(-\frac{e^{a t+1} (c-a k)}{c}\right)-1}{a}
\end{equation}
where $W$ is now the Lambert W-function. As a result, we are now in the position of obtaining $P(k,t)$, given by the eqn. (\ref{ansatz}):
\begin{equation}
P(k,t)=\frac{1}{c \left(-e^{W\left(\frac{e^{a t+1} (a k-c)}{c}\right)-a t-1}-1\right)+a k}
\label{eq:nopdist}
\end{equation}
which shows several interesting properties. It is easy to see that
\begin{equation}
P(k,t\gg1)\approx F(t) \frac{1}{k}.
\end{equation}
Thus we expect to observe, once one has normalized the distributions, a power law which is compatible with $k\approx 1$, once one has appropriately normalized the bins.

Two important comments are in order. First, we note that since $P(k,t)=P_1(k) P_2(t)$, one can, at each finite time $t$, obtain a distribution out of equilibrium for the degree which is consistently $P_1(k)\approx \frac{1}{k}$. However, it is easy to see that this distribution is not normalizable, as for large values of $k$, $\int P_1(k) dk \approx \log(k)$, which indeed suggests that this distribution cannot be a stable one.
Since $a<1$, being the product of the probability and the requirement that we remove only smaller values than $1$,  $\lim_{t\rightarrow \infty} P_1(t)=0$ and thus requires more statistical analysis as one goes further to obtain this distribution.  Thus, the distribution of degree is valid only for finite values of $k$, as one would expect, and has to be normalized by a factor $\tilde P_1(k)=\frac{1}{N}\frac{1}{k}$, with
\begin{equation}
N=\int_{\tilde k}^{k=t} \frac{1}{\xi} d\xi=\log(\frac{t}{\tilde k}).
\end{equation}
In addition, we note that unless the $k> \frac{c}{a}$, $P(k,t)$ does not exist, as the W-function is defined only for positive values.

\begin{figure}
\includegraphics[scale=0.5]{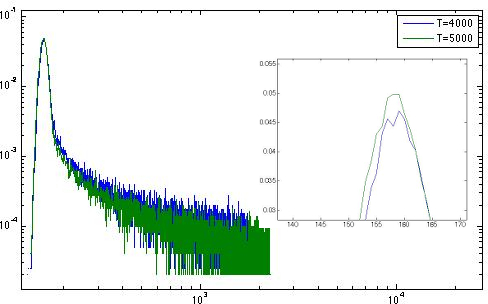}
\caption{Distribution of degree obtained for $p_{d}=p_{nl}=M=1$, $\alpha=0.1$ and no particles. The distributions are compared at two different times, $T=4000$ (blue) and $T=5000$ (green). We can observe that for the tail at later time is lower than the one at earlier time. This is compatible with the observation of the instability of the tail from eqn. (\ref{eq:nopdist}), as this is roughly 4/5 higher than the other one, and the probability is transferred to the lower degree bin (inset).}
\label{fig:instability}
\end{figure}

We have discussed the case of the probability distribution obtained by acceleration of the network in combination with decay, and have shown that the probability distribution is unstable, as it effectively decays as $1/t$, as explained in Fig. \ref{fig:instability}, and thus compatible with the simulations performed.

\begin{figure}[!htbp]
\includegraphics[scale=0.4]{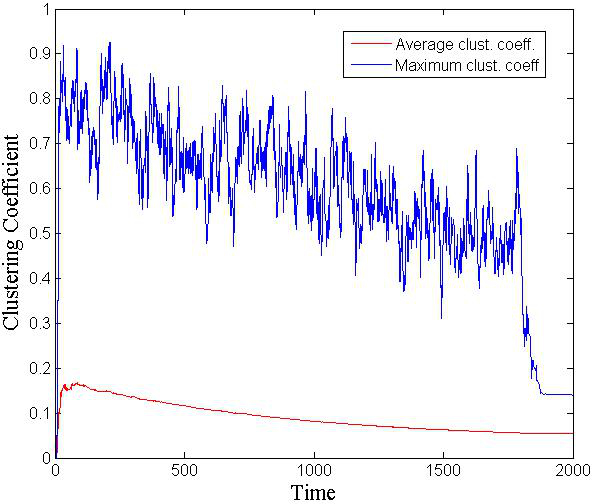}
\caption{Maximum and average clustering coefficient at  $T=2800$, $\gamma=\alpha=0.1$, $p_d=0.01$, $p_p=1$ and $L_d=0.99$, for which we obtained a degree distribution with a power law tail with exponent $\rho=-2.36$.
The plot shows that only few nodes are highly clustered, while the average is low. The drop at $T=2800$ is due to the fact that the growth stopped,
and we evolved the system (so that only stable nodes would survive). Results averaged over 30 simulations.}
\label{fig:clc}
\end{figure}

\textit{Clustering} In order to confirm we have indeed obtained a small-world network, we also studied the clustering coefficient.
This is defined as follows for weighted graphs. The number of triangles based at a node $i$, if $A_{ij}$ describes the network, is given by $t(i)=\sum_{jk} \frac{1}{2} A_{ij}A_{jk}A_{ki}$. The clustering coefficient $C(i)$ of a node $i$ is the ratio of number of triangles and the immediate weighted neighbours: $C(i)={2t(i)}/{\sum_{j=1}^N A_{ij}}$. We thus define $C_{max}=max_i C(i)$, while $C_{mean}=\frac{1}{N} \sum_{i=1}^N C(i)$. This is plotted in Fig. \ref{fig:clc}, for the parameters $p_p=1$, $p_d=0.01$, $\alpha=\gamma=0.1$, $L_d=0.99$, as in Fig. \ref{fig:clc}.
The maximum clustering coefficient $C_{max}$ of the graph is rather high, although it is clear that as the graph increases, $C_{max}$ decreases until saturation as shown in Fig.\ref{fig:clc}, and it drops as the growth stops (T=2800), while the average is basically unaffected, meaning that only a few nodes have high clustering coefficient.


\textit{Robustness.--- } The above discussion showed that there is an optimal memory range, as represented by the values of $M$, for the size of the tail. However, this by no means requires fine-tuning of that parameter. We show here that no fine-tuning is necessary in any of the other parameters of the model as well, namely the probability of decay $p_d$, the strength parameter $\gamma$, the probability of connecting a new node to an old one $p_{nl}$, and the stability threshold $L_d$. The results are shown in Figs. \ref{fig:changememory}, \ref{fig:changeparam}, \ref{fig:changedynlinkprob}, \ref{fig:threshvar} where it is clearly shown that the power-law degree distribution is robust with respect to variations in all  these parameters.

What is important, instead, is the continuous growth of the graph. Unless one imposes some geometry skeleton (as in \cite{ikeda}), if one stops the growth of the graph and keeps it evolving in time (therefore, decay and hopping/strengthening), the tail in the distribution is soon destroyed, depending on the thermalization time.

We previously mentioned that we observe tails both with exponent greater and lower than 2. We observe that the exponent depends linearly with the parameter which regulates the introduction of a new particle with each new note, the probability parameter $p_p$, such that $\rho\approx \rho_0 p_p$. However, a more detailed analysis is necessary in order to clearly find the dependence on the other parameters.


\end{document}